\documentclass[12pt,letter]{article}
\usepackage[margin=1in]{geometry}
\usepackage[colon]{natbib}
\usepackage{tabularx}
\usepackage{epsfig}
\usepackage[titletoc, title]{appendix}
\usepackage{setspace}
\usepackage{lscape}
\usepackage{booktabs}
\usepackage{array}
\usepackage{tabularx}
\usepackage[latin1]{inputenc}
\usepackage[english]{babel}
\usepackage{makeidx}
\usepackage{arydshln}
\usepackage{latexsym}
\usepackage{ifthen}
\usepackage{color,epsfig}
\usepackage{graphicx}
\usepackage{subfig}
\usepackage{mathrsfs}
\usepackage{wrapfig}
\usepackage{rotating}
\usepackage{natbib}
\usepackage{amsmath, amsthm}
\usepackage{nomencl}
\usepackage{float}
\usepackage{multirow}
\usepackage{multicol}
\usepackage{xcolor}
\usepackage{epic,eepic}
\usepackage{lscape}
\usepackage{tabularx}
\usepackage{pdflscape}
\usepackage{epstopdf}
\usepackage{calligra}
\usepackage{epsfig}
\usepackage[colon]{natbib}
\usepackage{tabularx}
\usepackage{amsmath}
\usepackage{amssymb}
\usepackage{pdflscape}
\usepackage{multirow}
\usepackage{multicol}
\usepackage[bottom]{footmisc}
\usepackage{array,arydshln}
\usepackage{appendix}

\newcolumntype{x}[1]{>{\centering\arraybackslash\hspace{0pt}}p{#1}}

\newcommand{\bs}{\boldsymbol}

\def\b#1{\mbox{\boldmath $#1$}}    

\begin{document}

\title{A non-homogeneous hidden Markov model for partially observed longitudinal responses}

\author{Maria Francesca Marino \\\texttt{mariafrancesca.marino@unifi.it} \and Marco Alf\'o\\ \texttt{marcoalfo@uniroma1.it}}

\date{}

\maketitle
\doublespace
\abstract{Dropout represents a typical issue to be addressed when dealing with longitudinal studies. If the mechanism leading to missing information is non-ignorable, inference based on the observed data only may be severely biased. A frequent strategy to obtain reliable parameter estimates is based on the use of individual-specific random coefficients that help capture sources of unobserved heterogeneity and, at the same time, define a reasonable structure of dependence between the longitudinal and the missing data process. We refer to elements in this class as random coefficient based dropout models (RCBDMs). 
We propose a dynamic, semi-parametric, version of the standard RCBDM to deal with discrete time to event. Time-varying random coefficients that evolve over time according to a non-homogeneous hidden Markov chain are considered to model dependence between longitudinal responses recorded from the same subject. A separate set of random coefficients is considered to model dependence between missing data indicators. Last, the joint distribution of the random coefficients in the two equations helps describe the dependence between the two processes.
To ensure model flexibility and avoid unverifiable assumptions, we leave the joint distribution of the random coefficients unspecified and  estimate it via nonparametric maximum likelihood. The proposal is applied to data from the Leiden 85+ study on the evolution of cognitive functioning in the elderly.

\vspace*{5mm}
 
}

\section{Introduction}
Missingness represents a frequent issue to be handle in longitudinal studies, as some participants may not be available at all intended time occasions and, therefore, may present incomplete data records. Monotone missingness represents the most frequent type of non-participation, with some individuals leaving the study prematurely and having a zero probability to re-enter. 
\cite{Rubin1976} introduced a well known taxonomy for missing data mechanisms which can be either based on the potential link between the longitudinal and the drop-out process or on the impact of missing data on parameter estimates in the longitudinal data model. In this respect, the drop-out may be non-ignorable in the sense that, even after conditioning on the observables (both covariates and responses), the participation to the study still depends on future (potentially unobserved) response values. Obviously, such a phenomenon may bias the study design and the resulting inference. 

Different modeling approaches to deal with such non-ignorable missingness may be found in the statistical literature. \cite{Little1995} and \cite{LittleRubin2002} distinguish two broad classes of models to account for missing responses: selection and-pattern mixture
models. Differences rely on the factorization of the joint distribution for the longitudinal and the missing data process which, in turn, may be linked to different interpretation of the dependence between the two processes. 
A further modeling alternative is based on the inclusion of individual-specific, typically Gaussian, random coefficients in the model specification to capture both the dependence between repeated measurements from the same individual (within-individual dependence) and between the longitudinal and the missing data process (between-outcomes dependence).
In the simplest case, the same set of random coefficients is shared by the two processes, with the resulting model being usually referred  to as {shared parameter model} \citep[e.g.][]{WuCarroll1988, WuBailey1988, DeGuttolaTu1994}.
Missing data models with nonparametric (discrete) random effects/coefficients were proposed by \cite{AlfoAitkin2000}, \cite{Roy2003},  \cite{Tsonaka2009}, and  \cite{AlfoMaruotti2009}.
As noticed by \cite{Bartolucci2015}, a limitation of standard shared parameter models is that random coefficients are assumed to be constant over time. This may be appropriate with dealing with short time series only.
For this reason, they proposed a shared parameter model for multivariate longitudinal responses and a (discrete) time to event, based on the use of time-varying and time-constant (discrete) random intercepts which are both shared by the longitudinal and the missingness model. 

However, an implicit assumption of shared random parameter models is that the primary outcome and the drop-out mechanism are influenced by the same sources of unobserved heterogeneity. Clearly, this assumption may be restrictive in some cases and, above all, cannot be verified by looking at the observed data only. A more general approach to deal with non-ignorable missingness is that of considering two separate sets of random coefficients in the equations for the longitudinal and the drop-out process. The corresponding joint distribution is used to describe the dependence between these latter; see, among others, \cite{AlfoMaruotti2009}, \cite{cre2010,cre2011}, \cite{Gottfredson2014}, and \cite{Barrett2015}. 
While a Gaussian multivariate distribution is frequently assumed also in this case, a nonparametric specification can be considered to improve model flexibility as in \cite{Spagnoli2017}. 
As with shared parameter models, the assumption of time-constant random coefficients may be restrictive in some cases  \citep{Bartolucci2015}. While a possible strategy to describe how outcomes evolve over time may consist in including in the model specification some function of time associated to fixed or random parameters, such an approach would help us describe ``well-shaped'' (e.g. polynomial) dynamics only. A more flexible and appealing approach  may be based on the use of a Hidden Markov Model (HMM) formulation \citep{Zucchini2009, BartFarcoPennoni2013}. 

In this respect, we introduce a {random coefficient based} hidden Markov model for longitudinal responses subject to possible non-ignorable dropouts. 
To describe the dependence \textit{within} profiles, that is between longitudinal responses and missingness indicators from the same subject, we exploit two different sets of random coefficients. 
For the longitudinal outcome, we consider time-varying (discrete) random coefficients that evolve over time according to a non-homogeneous hidden Markov chain. These allows us to capture differential dynamics in the longitudinal responses over time.
On the other hand, for the missing data indicator, we consider time-constant (discrete) random coefficients that allows us to identify differential propensities to stay into the study. 
Dependence \textit{between} the two profiles is modelled via an upper-level latent class variable, with unspecified distribution 
which is estimated in a nonparametric maximum likelihood framework.

Our proposal is applied to data from the Leiden 85+ study, where the effect of demographic and genetic factors on the evolution of cognitive functioning for the elderly represents the main target of inference. Due to poor health conditions or death, a number of individuals enrolled in the study present incomplete data sequences. Inference based on the observed data only may lead to biased parameter estimates as, due to the study design, dependence between the drop-out and the unobserved responses is quite reasonable. 

The paper is organized as follows. In Section \ref{sec_LeidenIntro}, we describe the Leiden 85+ study. In Section \ref{Random_coeff_based_mod}, we briefly introduce random coefficient based drop-out model, while in Section \ref{sec_dyRCBDM}  we describe the proposed model specification. Section \ref{sec_estimation} entails the EM algorithm for maximum likelihood parameter estimation and the procedure to derive the estimated standard errors. In Section \ref{sec_laiden_analysis}, we describe the results from the application  of the proposed approach to the Leiden 85+ data. The last section, contains concluding remarks.

\section{The Leiden 85+ study}\label{sec_LeidenIntro}
The Leiden 85+ study is a longitudinal study conducted by the Leiden University Medical Center in the Netherlands, with the aim of analysing the evolution of cognitive functioning in the elderly. The study entailed Leiden inhabitants who turned 85 years old between September $1997$ and September $1999$. Out of $705$ subjects who were eligible to the study, $14$ died before they could be enrolled, $92$ refused to participate into the study, and $38$ refused to provide blood sample. At the end, $561$ elderly were followed up to six consecutive yearly visits until $90$ years of age. See \cite{derWiel2002} and \cite{vanVliet2010} for further details on the study.
Participants' cognitive conditions were assessed via the Mini Mental Status Examination index \citep[MMSE, ][]{Folstein1975}, which is obtained by evaluating the attention, the orientation, the language skills and the ability of the participant to perform simple actions. The corresponding questionnaire is based on $30$ binary (sub-) items grouped into seven different cognitive areas. Interviewers assigned value $1$ to correct answers, so that the MMSE scores (defined as the sum over the $30$ sub-items) can take all integer values in the interval $[0,30]$.
The study aims at identifying demographic and genetic factors that influence cognitive functioning and healthy ageing. To this purpose, the following covariates  were measured at the study start: \textit{gender}, \textit{educational status} -- a binary variable equal to $1$ for not less than $7$ years of schooling, and \textit{APOE genotype} --  a categorical variable identifying the Apolipoprotein E genotype of the patient. The three largest genotype groups ($\epsilon_{2}, \epsilon_{3}, \epsilon_{4}$) and their products were considered, leading to four different categories for this variable: $APOE_{22-23}, APOE_{24}, APOE_{33}$, and $APOE_{34-44}$. APOE genotype is known to play some role in aging. In particular, $\epsilon_{4}$ allele is known to be linked to an increased risk for dementia, whereas $\epsilon_{2}$ allele carriers are relatively protected.

\begin{figure}[ht]
\caption{Leiden 85+ data: distribution of the MMSE scores at each visit}
\centering
\includegraphics[scale=0.35]{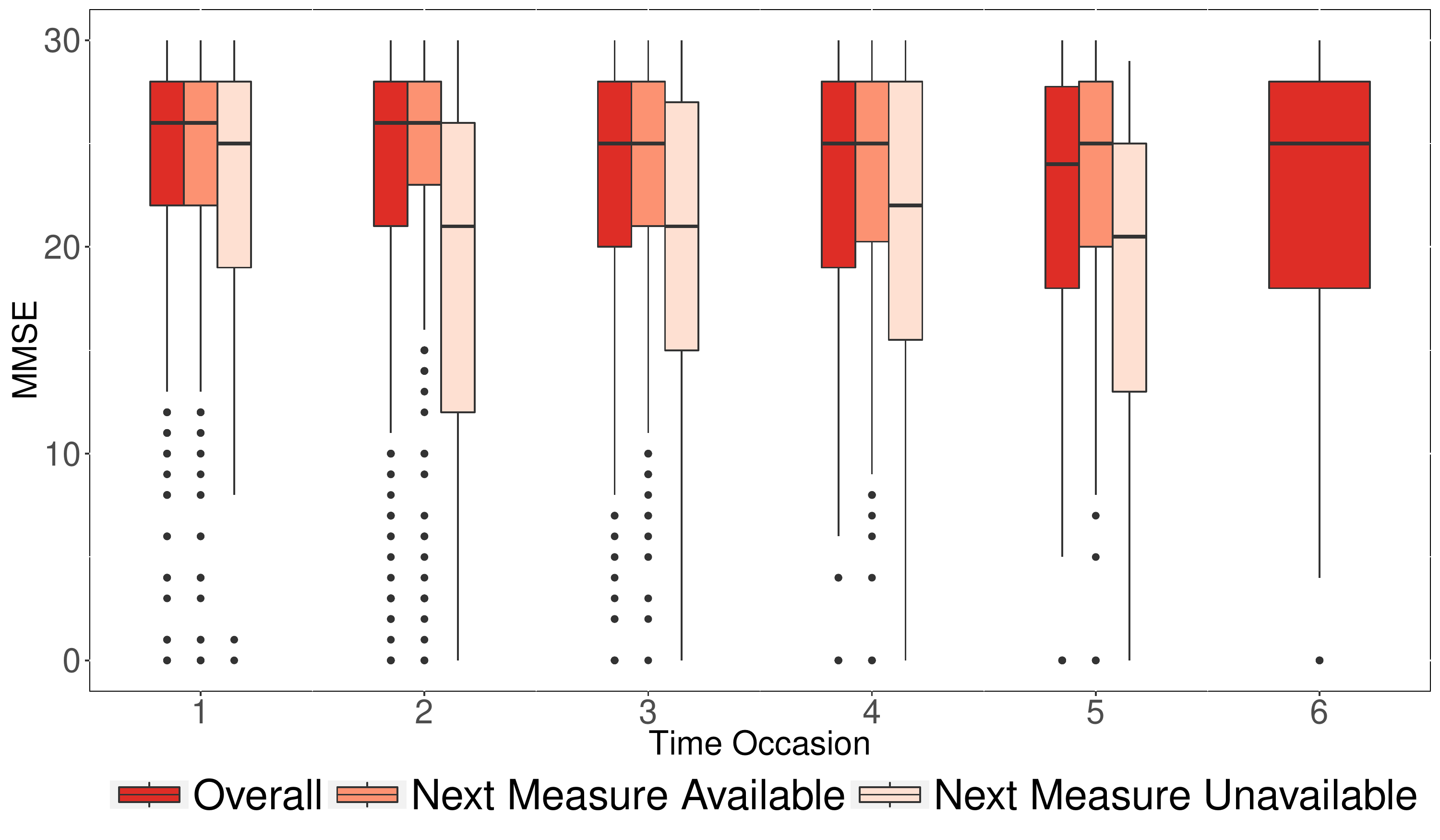}
\label{fig_meanResp}
\end{figure}

Due to the design of the study, a number of participants present incomplete responses (i.e. dropout), due to poor health conditions or death. We report in Figure \ref{fig_meanResp} the distribution of the MMSE scores at each visit stratified by whether subjects dropout between the current and the next occasion.
Based on this figure, it is clear that the overall response reduces with time; that is, as expected, cognitive skills reduce with people getting older. However, from this graph, it is also evident that MMSE  values tend to reduce faster for subjects dropping out prematurely. Furthermore, difference with those remaining longer under observation is more evident when the drop-out is observed at the beginning of the study. Such a finding poses the question on whether the process leading to missing data can be ignored.

\section{Random coefficient based drop-out model}\label{Random_coeff_based_mod} 

Let us suppose a longitudinal study is designed to collect measures for a response variable $Y_{it}, i = 1, \dots, n, t = 1, \dots, T,$ on a sample of $n$ individuals at $T$ time occasions. As it is frequent with longitudinal studies, some individuals in the sample may dropout prematurely and, thus,  present incomplete sequences. In this framework, let $\b R_i = (R_{i1}, \dots, R_{iT})^\prime$ denote the $T$-dimensional missing data vector, with $R_{it} = 0$ if the $i$-th subject is available at time occasion $t$ and $R_{it} = 1$ otherwise.  As we focus on monotone missingness, $R_{it} = 1 \Rightarrow R_{it^\prime} = 1, t, t^\prime = 1, \dots, T$, with $t^\prime > t.$
Let $\b b_{i} = (b_{i1}, \dots, b_{im})^\prime$ denote a vector of  individual-specific random coefficients in the longitudinal data model, with $ E(\b b_i) = \b 0$ and $\text{Cov}(\b b_i) = \b \Sigma_b$, $i=1,\dots,n$. As it is standard when dealing with random coefficients, we assume  that, conditional on $\b b_i$, longitudinal measures coming from the same subject are independent. Denoting by $\b Y_i = (Y_{i1}, \dots, Y_{iT})^\prime$ the vector of responses associated to the $i$-th subject in the sample, the corresponding joint (conditional) density is given by: 
\[
f_y(\b y_i \mid \b b_{i}) = \prod _{t = 1}^{T} f_y ( y_{it} \mid \b b_{i}).
\]
Also, let $\b c_{i} = (c_{i1}, \dots, c_{ir})^\prime$ denote a further set of individual-specific random coefficients that describe dependence in the missing data process, with $E(\b c_i) = \b 0$ and $\mbox{Cov}(\b c_i) = \b \Sigma_c.$  The notation is completed by defining $T_i^* = \min \, (T_i+1, T)$, where $T_i =  T - \sum_{t = 1}^T R_{it}$ denotes the number of available measures for individual $i = 1, \dots, n$. 
Conditional on $\b c_i$,  random variables $R_{i1}, \dots, R_{iT_i^*}$ are assumed to be independent, with joint (conditional) density given by: 
\[
f_r (\b r_{i} \mid \b c_i) = \prod_{t = 1}^{T^*} f_r (r_{it} \mid \b c_i).
\]

An essential feature of RCBDMs is the assumption of conditional independence between the longitudinal and the missing data process. That is, conditional on the individual-specific vectors $(\b b_i, \b c_i)$, $\b Y_i$ and $\b R_i$ are assumed to be independent.
In this framework, denoting by $\b Y_i^o$ and $\b Y_i^m$ the observed and the missing data in the individual sequence $\b Y_i = (\b Y_i^o, \b Y_i^m)$, the individual observed-data likelihood can be obtained as:
\begin{align}\label{ind_likelihood}
L_i(\cdot) &=  \int_{\mathcal B} \int_{\mathcal C}  \int_{\mathcal {Y}_i^m} f_y( \b y_{i}^o, \b y_{i}^m \mid \b b_{i})  f_r(\b r_{i} \mid\b c_i)  f_{b,c}(\b b_{i},\b c_i) \:d \b y_{i}^m \: d \b b_i \: d \b c_i,  \nonumber\\
& = 
\int_{\mathcal B} \int_{\mathcal C}  f_y(\b y_{i}^o\mid \b b_{i})  f_r(\b r_{i} \mid\b c_i)  f_{b,c}(\b b_{i},\b c_i) \: d \b b_i \: d \b c_i
\end{align}
where $f_y( \b y_{i}^o \mid \b b_{i}) = \prod_{t = 1}^{T_i} f_y( y_{it}\mid \b b_{i})$.

While it is common to assume that individual-specific random coefficients $(\b b_i, \b c_i)$ follow a specific parametric (usually Gaussian) distribution, a more flexible alternative recasts this problem in a finite mixture framework. 
Let $ Z_{i}$ denote an individual-specific latent variable defined on the support set $\{1, \dots, G\}$ with masses $\pi_g = \Pr(Z_{i} = g)$. 
In the $g$-th component of the finite mixture, random coefficients  $\b b_i$ and $\b c_i$ take value $\b \zeta_g$ and $\b \xi_g$, respectively, $g = 1, \dots, G$. 
In this context, the individual contribution to the observed data likelihood in equation \eqref{ind_likelihood} can be re-formulated as
\begin{equation}\label{fm_ind_likelihood}
L_i(\cdot) = \sum_{g=1}^G f_y( \b y_{i}^o \mid  Z_{i}=g) \: f_r (\b r_{i} \mid Z_{i}=g)\: \pi_g.
\end{equation}

Although the finite mixture representation is based on a robust and consistent (nonparametric) estimate of $f_{b,c}(\b b_{i},\b c_i)$, it is evident that this approach is not defined to (effectively) distinguish \textit{within} and \textit{between} profile dependence. This is due to the intrinsic {unidimensionality} of the discrete latent variable $Z_i.$ For this reason, \cite{Spagnoli2017} proposed a generalization of the model above, where two distinct sets of latent variables, with a possibly different number of categories, are considered. A first variable, $Z_{i} \in \{1, \dots, G\}$, is used to account for the dependence between longitudinal responses from the same subject, while a second one, $U_{i} \in \{1, \dots, K\}$, allows to model the dependence in the individual sequence of missing data indicators.  This formulation leads to the following expression for the individual contribution to the observed data likelihood:
\[
L_i(\cdot) = \sum_{g=1}^G \sum_{k=1}^K f( \b y_{i}^o \mid  Z_{i}=g) \: f (\b r_{i} \mid U_{i}=k)\: \pi_{gk},
\]
where $\pi_{gk} = \Pr(Z_i = g, U_i = k)$ is used to model dependence between $\b Y_i$ and $\b R_i$. Obviously, when $\pi_{gk}  = \pi_{g\cdot} \pi_{\cdot k}$, with $\pi_{g \cdot} = \sum_{k = 1}^K \pi_{gk}$ and $\pi_{\cdot k } = \sum_{g = 1}^G \pi_{gk}$, independence holds and maximum likelihood estimates for the longitudinal model parameters can be obtained by ignoring the missingness process.

\section{A dynamic representation}\label{sec_dyRCBDM}
In some cases, the hypothesis of time-constant random coefficients can be too restrictive for the longitudinal outcome and may not help model individual-specific latent dynamics \citep{Bartolucci2015}. In this section, we introduce a dynamic specification of the RCBDM described in the previous section. As we have limited information on $R_{it}$, we will consider time-varying random coefficients for the longitudinal data model only. The proposal can be easily generalized to deal with generic dynamic random coefficients when needed, e.g. when modeling the time to drop-out ($T_i$) via a discrete time (parametric) survival model.

Let $Z_{it}$ denote an individual-specific, time-varying, latent variable defined on the support set $\{1, \dots, G\}$ and let $\b Z_{i} = (Z_{i1}, \dots, Z_{iT_i})$ be the $T_i$-dimensional latent vector associated to the $i$-th subject, $i = 1, \dots, n$. In the following, $z_{it}$ and $\b z_{i}$ will be used to denote the generic realizations of $Z_{it}$ and $\b Z_{i}$, respectively.
As before, $U_{i}$ denotes an individual-specific, time-constant, latent variable defined on the support set $\{1, \dots, K\}$, while $u_{i}$ is the corresponding realization, with $i = 1,\dots, n$. We assume that latent variables $\b Z_i$ and $U_i$ influence the longitudinal and the missing data process, respectively. 
In particular, conditional on $\b Z_{i}=\b z_i$, the generic element $Y_{it}$ of the longitudinal vector $\b Y_i$ depends on $Z_{it}$ only and the joint (conditional) density for the observed longitudinal sequence from the $i$-th subject is 
\[
f_y(\b y_i^o \mid \b Z_i = \b z_i) = \prod _{t = 1}^{T_i} f_y( y_{it} \mid Z_{it} = z_{it}).
\]
We further assume that, conditional on $U_{i}=u_i$, the binary random variables $R_{i1}, \dots, R_{iT_i^*}$ are independent with joint (conditional) density 
\[
f_r(\b r_i \mid U_i = u_i) = \prod _{t = 1}^{T_i^\star} f_r( r_{it} \mid U_i = u_{i}).
\]
To describe the effect of the observed covariates on the outcomes $Y_{it}$ and $R_{it}$, the following regression models are also defined:
\begin{align}
\left\{ 
\begin{array}{l}
g[\text E(Y_{it} \mid Z_{it} = g)] ={ \zeta}_g + \b x_{it} ^ \prime \boldsymbol \beta, \\
\text{logit}[\Pr(R_{it}= 0 \mid  U _{i} = k)] = { \xi}_k + \b w_{it}^\prime \bs \gamma.
\end{array}
\right.
\end{align}
In the expressions above, $g(\cdot)$ represents an appropriate link function, while the parameters $ \bs \beta$ and $\bs \gamma$ denote the fixed effects associated of covariates $\b x_{it}$ and $\b w_{it}$, respectively.
Last, $ \zeta_g$ and $\xi_k$ denote the discrete random intercepts in the longitudinal and in the missing data model associated to $Z_{it} = g$ and $U_i = k$, respectively, for 
$g = 1, \dots, G$ and $k = 1, \dots, K$. 

To model the potential dependence between $\b Z_i$ and $U_i$ and, therefore, between the longitudinal and the missing data process, let us consider a discrete \textit{upper-level} latent variable $V_{i}$ defined on the support set $\{1, \dots, H\}$, with $\tau_h = \Pr(V_{i} = h)$ for $h = 1, \dots, H$. 
We assume that, conditional on $V_{i}$, the  latent variables $\b Z_i$ and  $U_i$ are independent, with joint distribution described by the following mixture model:
\begin{align*}
f_{z,u}(\b Z_i, U_i) &= \sum_{h = 1}^H  \tau_h \left[ \Pr(\b Z_i = \b z_i \mid V_i = h) \, \Pr(U_i = u_i\mid V_i = h)\right].
\end{align*}

In particular, we assumed that,  conditional on the $h$-th component of the upper-level mixture, that is $V_i = h$, the latent variables $Z_{it}$ evolve over time according to a first order hidden Markov chain, with initial probability vector $\bs \delta_h$ and transition probability matrix $\b Q_h$. The corresponding elements are given by
\begin{align*}
\delta_{g\mid h} &= \Pr({Z}_{i1} = g \mid V_i = h), \\
q_{gg^\prime \mid h} &= \Pr({Z}_{it} = g^\prime \mid {Z}_{it-1} = g, V_i = h),
\end{align*}
with $g, g^\prime = 1, \dots, G$ and $h = 1, \dots, H$.
As it can be noticed, the adopted parameterization is quite complex and, this could lead to an over-parameterized model. Therefore, in order to avoid numerical difficulties in deriving maximum likelihood estimates
and to reduce the number of parameters, we define $\b \delta_h$ and $\b Q_{h}$  according to the global logit parameterization suggested by \cite{ColombiForcina2001}.

To start, let us introduce the following set of inequality constraints on the random intercepts in the longitudinal data process: 
\begin{align}
\label{zeta_constr}
\zeta_1 \leq \zeta_2 \leq \dots \leq \zeta_G,
\end{align}
so that lower values of $\zeta$ (and therefore of $Z_{it}$) correspond to lower expected values for the longitudinal responses. 
Furthermore, we define the initial probabilities $\delta_{g\mid h}$ as follows:
\begin{align}\label{eq_init}
\log \frac{\Pr(Z_{it}\geq g \mid V_i = h)} {\Pr(Z_{it}< g \mid V_i = h)} 
= \log \frac{\delta_{g\mid h} + \dots \delta_{G\mid h}} {\delta_{1\mid  h} + \dots + \delta_{g-1 \mid h}} 
= \alpha_{0g} +\psi_{0h},
\end{align}
with $ h = 1, \dots, H$ and $g=2,\dots,G$. For identifiability purposes, we set $\psi_{01}=0$, so that the number of parameters to be estimated reduces from $H(G-1)$ to $(G-1)+(H-1)$.
Similarly, the transition probabilities $q_{g g^\prime\mid h}$ are modelled as follows:
\begin{align}
\log \frac{\Pr(Z_{it}\geq g \mid \Pr(Z_{it-1}=g^\prime, V_i = h)} {\Pr(Z_{it}< g \mid \Pr(Z_{it-1}=g^\prime, V_i = h)}  = 
\log \frac{q_{gg^\prime\mid h}+ \dots + q_{Gg^\prime \mid h}}{q_{1g^\prime\mid h} + \dots + q_{g-1 g^\prime\mid h}} & = \alpha_{1g g^\prime} + \psi_{1h}, \label{eq_Q}
\end{align}
with $h = 1,\dots, H$, $g= 1, \dots, G$, and $g^\prime = 2, \dots, G$. As above, to ensure parameter identifiability, we set $\psi_{11} = 0$, so that $G(G-1)+(H-1)$ parameters need to be estimated rather than $HG(G-1)$.

\subsection{Model interpretation}\label{sec_interpretation}

The modeling approach we propose offers a great flexibility. To start, unverifiable parametric assumptions on the random coefficient distribution may be avoided. Also, different individual-specific behaviors may be accommodated via this bi-dimensional latent structure. In particular, the upper-level latent variable influences both the way subjects move across the states of the hidden Markov chain and their propensity to stay into the study. Transitions between states may be more relevant for subjects belonging to a specific upper-level class and less relevant for others. Also, different propensities to dropout from the study described by the latent variable $U_i$ may be observed for different upper-level classes. 
For instance, as we will see for the analysis of the Leiden 85+ data, the upper-level latent variable $V_i$ may help identify different health conditions of subjects under observation. In the set of subjects with a better (worse) condition, we may distinguish those with a higher propensity to drop-out from the study, those with a lower propensity to drop-out, and those with complete data records. These differences may be described by the latent variable $U_i$ in the missing data model. In a similar fashion, the health conditions of a subject may lead to a different evolution in the longitudinal response over time. By letting the hidden Markov chain $\b Z_i$ depend on the upper-level latent variable $V_i$, non-homogeneous dynamics in the longitudinal profiles can be easily accommodated. 

A further advantage of the proposed model specification is related to the chance of accounting for a potentially non-ignorable missing data process when $H>1$.
The independence model, leading to an ignorable missingnes,s is directly nested within the proposed parameterization and corresponds to $H = 1$. In this case, the joint density $f_{z,u}(\b Z_i, U_i)$ factorizes into the product of the corresponding marginals, i.e. $f_{z,u}(\b Z_i, U_i) =  f_z(\b Z_i)  f_u( U_i),$ and the individual sequences $\b Y_i$ and $\b R_i$ are independent. Therefore, in a sensitivity analysis perspective, {$H$ can also be interpreted as a non-ignorability parameter for the missing data process. }

\section{Parameter estimation} \label{sec_estimation}
Estimation of model parameters can be carried out using a maximum likelihood approach. Due to the local independence assumption, the observed data likelihood is defined by
\begin{align*}
L(\b \theta) &= 
\prod_{i = 1}^n \sum_{h}^H \tau_h 	\: \left\{\sum_{{z}_{i1}\cdots {z}_{iT_i} }
\left[
 \prod_{t = 1}^{T_i}
f_y( y_{it} \mid  Z_{it} = z_{it})  \, 
\delta_{{z}_{i1}\mid h}  \, \prod_{t = 2}^T q_{ {z}_{it-1} {z}_{it}\mid h} \right] \right. \times \\
& \left. \times \left[ 
\prod_{t = 1}^{T_i^*} \sum_{u_i}
f_r(r_{it} \mid   U_i = u_{i})\, 
\pi_{u_i \mid h}
\right]  \right\},
\end{align*}
where $\b \theta$ denotes the set of all free model parameters and $\pi_{{u_i} \mid h}$ is used in place of $ \Pr(U_i = u_i\mid V_i = h)$ to simplify the notation.
To avoid multiple summations over all possible realizations of the hidden chain, $\b {Z}_{i}$, we may rely on the EM algorithm \citep{Dempster1977}. 
To this purpose, let $a_{itg}=1$ if the $i$-th subject is in the $g$-th state at occasion $t$, and let $a_{itgg^\prime} = a_{it-1g} \times a_{itg^\prime}$, with $g = 1, \dots, G$. Similarly, let $d_{ik}$ and $e_{ih}$ be the indicator variables for $U_i = k$ and $V_i = h$, respectively. 
To derive parameter estimates, we define the following complete data log-likelihood function: 
\begin{align}\label{completeL}
\small
\ell_c (\b \theta) &= 
\sum_{i= 1}^n  \left\{
\left[ \sum_{h=1}^H e_{ih} \log \tau_h\right] + 
\left[\sum_{h=1}^H\sum_{g = 1}^G  e_{ih} \,   a_{i1g} \log \delta_{g\mid h} +\right. \right.\nonumber   \\  
& \quad + \left. \left. \sum_{h = 1}^H\sum_{t = 2}^{T_i} \sum_{g = 1}^G \sum_{g^\prime = 1}^G  e_{ih} \,   a_{itgg^\prime} \log q_{g g^\prime\mid h} 
+\sum_{t = 1}^{T_i} \sum_{g =1}^G   a_{itg} \log f_y (y_{it}\mid {Z}_{it} = g)\right]\right. \nonumber \\
& \quad \left.  +\left[ \sum_{h = 1}^H \sum_{k = 1}^K  e_{ih} \,   d_{ik} \log \pi_{k \mid h}\right]  +\left[\sum_{t = 1}^{T_i^\star} \sum_{k = 1}^K    d_{ik} \log f_r(r_{it} \mid U_{i} = k)
\right] \right\}.
\end{align}

The E-step of the algorithm consists in calculating the expected value of the complete data log-likelihood over the unobserved component indicators, conditional on the observed data $(\b y_i^o, \b r_i)$ and the current value of the parameter estimates $\hat{\b \theta}^{(r)}$. That is, 
\begin{align}
\label{eq_expCompl}
\small 
 Q \left(\bs \theta \mid \hat{\bs \theta}^{(r)}\right) & = 
\sum_{i= 1}^n  \left\{
\left[ \sum_{h=1}^H \hat e_{ih} \log \tau_h\right] + 
\left[\sum_{h=1}^H\sum_{g = 1}^G \hat e_{ih} \,  \hat a_{i1g\mid h} \log \delta_{g\mid h} \right. \right. \nonumber \\  
& + \left. \left. \sum_{h = 1}^H\sum_{t = 2}^{T_i} \sum_{g = 1}^G \sum_{g^\prime = 1}^G \hat e_{ih} \, \hat  a_{itgg^\prime\mid h} \log q_{g g^\prime\mid h} 
+\sum_{t = 1}^{T_i} \sum_{g =1}^G  \hat a_{itg} \log f _y(y_{it}\mid {Z}_{it} = g)\right]\right. \nonumber \\
& \left.  +\left[ \sum_{h = 1}^H \sum_{k = 1}^K \hat e_{ih} \,  \hat d_{ik\mid h} \log \pi_{k \mid h}\right]  +\left[\sum_{t = 1}^{T_i^\star} \sum_{k = 1}^K   \hat d_{ik} \log f_r(r_{it} \mid U_{i} = k)
\right] \right\},
\end{align}
where indicator variables in equation \eqref{completeL} are replaced by the corresponding conditional expectations.
Such a computation can be consistently simplified by slightly modifying the standard forward/backward variable approach which is typically used in the hidden Markov model framework \citep{Baum1970, Welch2003}. To this purpose, let us define the forward and the backward variables as follows:
\begin{align*}
\mathcal A_{ig\mid h}^{(t)} &= f(y_{i1}, \dots, y_{it}, {Z}_{it} = g \mid V_{i} = h), \\
\mathcal B_{ig\mid h}^{(t)} &=
f ( y_{it+1}, \dots, y_{T_i} \mid {Z}_{it} = g, V_{i} = h).
\end{align*}
The above quantities can be recursively derived following similar arguments to those detailed in \cite{Baum1970}.
Once they are computed, the posterior expectation of the indicator variables in equation \eqref{completeL} is given by
\begin{align*}
 \hat a_{itg\mid h} &= 
\frac{\sum_{k}
\mathcal A_{ig\mid h}^{(t)}  \,  B_{ig\mid h}^{(t)} \, f_r(\b r_i \mid U_i = k)\,  \pi_{k\mid h}\,  \tau_h}
{\sum_g \sum_{k}
\mathcal A_{ig\mid h}^{(t)}  \,  B_{ig\mid h}^{(t)} \, f_r(\b r_i \mid U_i = k)\,  \pi_{k\mid h} \,  \tau_h},\\[0.2cm]
\hat a_{itgg^\prime \mid h} &= \frac
{\sum_{k}
\mathcal  A_{ig\mid h}^{(t-1)} \,  q_{g g^\prime \mid h} \,  f (y_{it} \mid \mathcal{Z}_{it} = g^\prime) \, \mathcal B_{ig^\prime \mid h}^{(t)} \, f_r(\b r_i \mid U_i = k)\,  \pi_{k\mid h} \,  \tau_h}
{\sum_{k} \sum_{g} \sum_{g^\prime}
A_{ig\mid h}^{(t-1)} \,  q_{g g^\prime \mid h} \,  f(y_{it} \mid \mathcal{Z}_{it} = g^\prime) \, \mathcal B_{ig^\prime \mid h}^{(t)} \,  f_r(\b r_i \mid U_i = k)\, \pi_{k\mid h} \,  \tau_h}, 		\\[0.2cm]
\hat d_{ik \mid h}&= \frac
{\sum_{g}
\mathcal A_{ig\mid h}^{(T_i)} \, f_r(\b r_i \mid U_i = k)\, \pi_{k\mid h} \, \tau_h}
{\sum_{k}\sum_{g}
  \mathcal A_{ig\mid h}^{(T_i-1)} \, f_r(\b r_i \mid U_i = k) \, \pi_{k\mid h} \, \tau_h}, 	 \\[0.2cm]
\hat e_{ih} &= \frac
{
\sum_{g}\sum_{k}
 \mathcal A_{ig\mid h}^{(T_i)} \, f_r(\b r_i \mid U_i = k) \,  \pi_{k\mid h} \, \tau_h
}
{
\sum_{ h}\sum_{g}\sum_{k}
\mathcal A_{ig\mid h}^{(T_i)} \, f_r(\b r_i \mid U_i = k) \,  \pi_{k\mid h} \, \tau_h
}.
\end{align*}
The remaining posterior probabilities can be computed as $\hat a_{itg} = \sum_h \hat a_{itg \mid h} \, \tau_h$ and $\hat d_{ik} = \sum_h \hat d_{ik \mid h}\,  \tau_h$.

In the M-step of the algorithm, we maximize equation \eqref{eq_expCompl} with respect to model parameters $\b \theta$. Due to the separability of the parameter space, we can partition the maximization into distinct sub-problems. 
Starting from expression \eqref{eq_expCompl}, it is easy to notice that the estimates for $\hat\pi_{k \mid h}$ and $\tau_h$ are given by
\[
\hat\pi_{k \mid h} = \frac{\sum_{i=1}^n \hat e_{ih} \hat d_{ik\mid h}}{\sum_{i=1}^n \sum_{k=1}^{K}\hat e_{ih} \hat d_{ik \mid h} }
, \quad \quad 
\hat \tau_h = \frac{1}{n}\sum_{i=1}^n \hat e_{ih},
\]
with $ k = 1, \dots, K$ and $h = 1, \dots, H$, respectively.  
To estimate the parameters for the hidden Markov chain, the following M-step equations need to be solved: 
\begin{align*}
&\sum_{i = 1}^n \sum_{h=1}^H\sum_{g = 1}^G \hat e_{ih} \,  \hat a_{i1g \mid h} \,  \frac{\partial \log \delta_{g\mid h}}{\partial \b \eta_0}
= \b 0, \\
&
\sum_{i = 1}^n \sum_{h=1}^H\sum_{g = 1}^G \sum_{g^\prime = 1}^G\hat e_{ih} \,  \hat a_{itgg^\prime \mid h} \,  \frac{\partial \log q_{gg^\prime\mid h}}{\partial \b \eta_1} = \b 0, 
\end{align*}
where $\b \eta_0 = \{\alpha_{0g}, \psi_{0h}, g = 2, \dots, G, h = 2, \dots, H\}$ and 
$\b \eta_1 = \{\alpha_{1gg^\prime}, \psi_{1h}, g = 1, \dots, G, g^\prime= 2, \dots, G, h = 2, \dots, H\}$.

The parameters for the longitudinal data process, $\b \Psi = ( \b \beta,  \zeta_1, \dots, \zeta_G)$, are updated by solving 
\[
\sum_{i=1}^n \sum_{t = 1}^{T_i} \sum_{g =1}^G  \hat a_{itg} \frac{\partial \log f(y_{it}\mid {Z_{it}} = g)}{\partial \b \Psi} = \b 0
\]
under the constraints: $\zeta_1 \leq \dots \leq \zeta_G$. 
Last, for the missing data process, we need to solve: 
\[
\sum_{i=1}^n  \sum_{t = 1}^{T_i}\sum_{k =1}^K  \hat d_{ik}  \frac{\partial \log f(r_{it}\mid U_{i} = k)}{\partial \b \Phi} = \b 0,
\]
where $\b \Phi = (\b \gamma,  \xi_1, \dots, \xi_K)$. 
The E- and the M-steps are iterated until convergence, specified in terms of the log-likelihood or the parameter values, using appropriate relative or absolute norms, e.g. $\| (\b \theta^{(r)} -  \ell(\b \theta^{(r-1)})\|< \varepsilon$ or $\| \b \theta^{(r)} -  \b \theta^{(r-1)}\| < \varepsilon$. 
To avoid local maxima, for a given choice of $(G,K,H)$, the algorithm is initialized from multiple starting values; at the end, the model with the highest log-likelihood value, is kept as the optimal solution.

\subsection{Standard errors and model selection}
Standard errors for parameter estimates obtained at convergence of the EM algorithm, $\hat{\b \theta}$, can be computed using the standard sandwich formula \citep[see e.g.][]{White1980, Royall1986}. 
For this purpose, we start by re-parameterizing some of the elements in $\b \theta $ to obtain a vector of unconstrained parameters via the following logit transforms:
\begin{align*}
\tau_h^* &= \log \frac{\tau_h}{\tau_1}, \quad h = 2,\dots,  H, \\
\pi_{k\mid h}^* & = \log \frac{\pi_{k \mid h}}{\pi_{1 \mid h}}, \quad k =2, \dots K, \quad h =1, \dots, H.
\end{align*}
Denoting by $\b \theta^*$ the vector of transformed parameter estimates, the sandwich estimate of the covariance matrix for $\b \theta^*$ is defined by
\begin{equation}
\label{eq_sw}
\widehat{\mbox{Cov}}(\hat{\b \theta}^* )  = {\b J}(\hat{\b \theta}^* )^{-1} \hat{\b K}(\hat{\b \theta}^*) {\b J}(\hat{\b \theta}^* )^{-1},
\end{equation}
where ${\b J}(\hat{\b \theta}^* )$ represents the observed information matrix and
$\hat{\b K}(\hat{\b \theta}^*)$ provides an estimate of the covariance matrix for the score vector, defined by
\[
\b K (\b \theta^*)=  \mbox{Cov} \left[ \frac{\partial \ell(\b \theta^*)}{\partial \b \theta^*} \right]=\sum_{i=1}^n \mbox{Cov} \left[ \frac{\partial \ell_i(\b \theta^*)}{\partial \b \theta^*} \right].
\]
In particular, ${\b J}(\hat{\b \theta}^* )$ is obtained by computing the first numerical derivative of the score vector $S({\b \theta}^*) = \partial \ell(\b \theta^*)/\partial \b \theta^*$, evaluated at $\hat{\b \theta}^*$. On the other hand,  $\b K (\b \theta^*)$ is estimated by $\hat{\b K}(\hat{\b \theta}^* ) = \sum_{i = 1}^n S_i(\hat {\b \theta}^*)S(\hat {\b \theta}^*)^\prime$, where $S_i(\hat {\b \theta}^*)$ denotes the individual contribution to the score function for the $i$-th subject, evaluated at ${\hat{\b \theta}}^*.$

Standard errors for $\hat{\b \theta}^*$ are obtained as the square root of the diagonal elements in $\widehat{\mbox{Cov}}(\b \theta^* )$. These can be expressed on the original scale by adopting the delta method:
\[
\widehat{\mbox{Cov}}(\hat{\b \theta}) = {\b M}(\b \theta^* ) \, \widehat{\mbox{Cov}}(\hat{\b \theta}^*) {\b M}(\b \theta^* ) ^\prime,
\]
where ${\b M}(\b \theta^* )  = \partial \b \theta / \partial (\b \theta^*)^\prime $ and, then, taking the values on the corresponding diagonal.

As it is frequent in the mixture model framework, the EM algorithm is run by treating the number of classes and states as fixed and known. The algorithm is run for varying choices for $(G, K, H)$ and the best model is chosen via model selection techniques (e.g. AIC - \citealp{Akaike1973} or BIC - \citealp{Schwarz1978})

\section{Back to the Leiden 85+ study}\label{sec_laiden_analysis}
In this section, we apply the dynamic RCBDM to the longitudinal data from the Leiden 85+ study we described in Section \ref{sec_LeidenIntro}. As we highlighted before, we aim at understanding the effect of demographic and genetic conditions on the evolution of cognitive functioning in the elderly. 
We expect the non-homogeneous hidden Markov chain to offer a clear description of the observed longitudinal sequences and to effectively capture the dependence between longitudinal responses coming from the same subject. 
When compared to  time-constant random coefficient models, HMMs turn out to be more flexible and provide a more concise description of the data.
Similarly, the non-homogeneous mixture for the drop-out model is expected to identify different individual propensities to stay under observation and capture dependence between missingness indicators.
Last, we expect the upper-level mixture to represent a flexible and effective way to describe the potential dependence between the longitudinal responses and the missing data indicators recorded from the same subject. Since the basic independence model ($H = 1$) is nested into the proposed specification, we may easily analyze the robustness of the estimates with respect to assumptions upon the dependence between the longitudinal response and the missingness indicators.
With these objectives in mind, we start from the definition of the following models for the longitudinal response and the missing data indicator: 
\[
\left\{ 
\begin{array}{l}
\text E(Y_{it} \mid  Z_{it} = g) = {\zeta}_g +\textbf x_{it} ^ \prime \boldsymbol \beta  \\
\text{logit}[\Pr(R_{it}= 0 \mid  U _{i} = k)] = { \xi}_k +\b w_{it}^\prime \bs \gamma  ,
\end{array}
\right.
\] 
where $Y_{it} = \log[1+ (30-\mbox{MMSE}_{it})]$ denotes the longitudinal response variable, while $\b x_{it}$ and $\b w_{it}$ represent the set of covariates in the two equations, respectively. 
In particular, the covariate set is common, that is $\b x_{it} = \b w_{it}$, and include age (measured in terms of deviation from the age at the study entry -- $85$), sex (reference = female), educational status (reference = primary), and APOE genotype (reference = $\mbox{APOE}_{33}$).
It is worth to highlight that, out of the $561$ subjects participating into the study, only $541$ provided complete covariate information, so that the analysis is conducted on these subjects only ($i = 1, \dots, 541$). 

To identify the optimal number of latent classes and states, we run the EM algorithm for $G, K = 2, \dots, 5$ and $H = 1,2,3$, considering a multi-start strategy based on $50$ starting values.
As it is frequently done in the HMM framework, the optimal model was selected according to the BIC index using the number of observed individuals $n$ to penalize the log-likelihood function. It is known that this is quite a conservative choice but, since model interpretability  is of primary interest, this represents a reasonable choice. Values of the adopted penalized likelihood criterion are reported in Table \ref{tab_leidenBic}.
\begin{sidewaystable}
\caption{Leiden 85+ data: model selection. BIC values for varying $H, G$ and $K$}
\label{tab_leidenBic}
\centering
\begin{tabular}{lcccccccccccccccccc}
\toprule
\toprule
&&\multicolumn{4}{c}{$H = 1$}&&\multicolumn{4}{c}{$H = 2$}&&\multicolumn{4}{c}{$H = 3$}\\
\cline{3-18}
&$K$&\multicolumn{1}{c}{2}&\multicolumn{1}{c}{3}&\multicolumn{1}{c}{4}&\multicolumn{1}{c}{5}&&\multicolumn{1}{c}{2}&\multicolumn{1}{c}{3}&\multicolumn{1}{c}{4}&\multicolumn{1}{c}{5}&&\multicolumn{1}{c}{2}&\multicolumn{1}{c}{3}&\multicolumn{1}{c}{4}&\multicolumn{1}{c}{5}\\
\midrule
\multirow{4}{*}{$G$}&
2 & 6175.95 & 6159.68 & 6162.46 & 6157.82 && 6183.51 & 6170.66 & 6175.04 & 6175.59 && 6208.19 & 6201.41 & 6212.03 & 6216.82 \\ 
&3 & 5701.23 & 5684.96 & 5687.74 & 5683.10 && 5693.74 & 5679.53 & 5683.69 & 5683.30 && 5717.47 & 5706.25 & 5713.48 & 5721.03 \\ 
&4 & 5491.87 & 5475.60 & 5478.38 & 5473.74 && 5480.68 & 5466.31 & 5469.13 & 5463.99 && 5495.70 & 5487.58 & 5495.00 & 5497.70 \\ 
&5 & 5289.48 & 5278.05 & 5280.83 & 5293.42 && 5269.38 & \textbf{5256.45} & 5261.76 & 5262.26 && 5276.92 & 5270.26 & 5278.42 & 5287.92 \\ 
  
\bottomrule
\bottomrule
\end{tabular}
\end{sidewaystable}
Based on these results, we may conclude that the optimal solution corresponds $H = 2$, $G=5$ and $K = 3$ ($\mbox{BIC} = 5256.45$). 
Selecting the model with $G = 5$ hidden states, we consider a solution which lies on the boundary of Table \ref{tab_leidenBic}; however, we decided to retain it, without proceeding further with higher $G$ values, to preserve the interpretability of model parameters.

\subsection{Results: the upper-level mixture}
\label{sec_LeidenUpper}
As stated before, the upper-level stratum of the proposed model specification allows us to describe unobserved, individual-specific, features determining the potential relation between the longitudinal and the missing data process. The selection of a model with $H =2 $ upper-level components defines a Missing Not At Random (MNAR) mechanism and suggests a potential dependence between the random effects in two equations. In particular, based on the estimated parameters, we may conclude that $21.9\%$ of the observed subjects belongs to the first upper-level class ($\tau_1 = 0.219$), while the remaining $78.1\%$ belongs to the second one ($\tau_2 = 0.781$). Obviously, they could be more easily interpreted if we look at the (conditional) behavior of individuals with respect to the longitudinal and the missingness process.

\subsection{Results: the longitudinal data process}
\label{sec_LeidenLong}
Parameter estimates for the longitudinal data model, with the corresponding standard errors, are reported in Table \ref{tab_leidenEst}.
\begin{table}[!h]
\caption{Leiden 85+ data: parameter estimates and standard errors for the longitudinal data model}
\label{tab_leidenEst}

\centering

\begin{tabular}{lrrrrr}
\toprule
\toprule
 & \multicolumn{1}{c}{Estimates} & \multicolumn{1}{c}{Se} \\ 
\midrule
$\zeta_1$ & 0.260 & 0.067 \\ 
$\zeta_2$ & 1.134 & 0.057 \\ 
$\zeta_3$ & 1.783 & 0.049 \\ 
$\zeta_4$ & 2.441 & 0.054 \\ 
$\zeta_5$ & 3.036 & 0.070  \\
\hdashline 
$Age$ & 0.056 & 0.007 \\ 
\textit{High Edu} & -0.265 &  0.055 \\ 
$Male$ & -0.100 & 0.055 \\ 
$\mbox{APOE}_{22-23}$ & 0.057 & 0.051 \\ 
$\mbox{APOE}_{24}$& -0.224 & 0.049 \\ 
$\mbox{APOE}_{34-44}$& 0.249 & 0.064 \\ 
\bottomrule
\bottomrule
\end{tabular}
\end{table}
Focusing on the estimated random intercepts, we may conclude that higher hidden states correspond to higher baseline response values, that is to individuals with worse cognitive functioning. 
As regards the covariate effects, estimates highlight that health conditions worsen with increasing age, while participants with higher education tend to be less cognitively impaired. This effect my be due to higher socio-economic states and to better life conditions especially during childhood. When looking at the effect of Apolipoprotein E genotype, a differential effect of $APOE_{24}$ and $APOE_{34-44}$ with respect to the baseline category ($APOE_{33}$) is observed. However, it is worth to highlight that the estimated parameters for $APOE_{24}$ should be carefully considered as the sample is strongly unbalanced with respect to alleles, and only $12$ subjects (out of $541$) present such a condition.

To better understand how cognitive functioning evolves over time, we may look at the estimated parameters for the hidden Markov chain. 
These suggest how subjects in the $h$-th upper-level class move across hidden states over time and, therefore, we may get a clearer description of the observed longitudinal patterns.
With the aim of reducing the number of estimated parameters, we considered the parametric specifications in equations \eqref{eq_init} and \eqref{eq_Q} for $\b \delta_h$ and $\b Q_h$, with $h = 1, 2$. Such specifications lead to the results reported in Table \ref{tab_leiden_hmm}.
\begin{table}[!h]
\caption{Leiden 85+ data - longitudinal data model: initial $(\b \delta_h)$ and transition ($\b Q_h$) probabilities by upper-level components}
\label{tab_leiden_hmm}
\centering
\begin{tabular}{lcccccc}
\toprule
\toprule
 & State & 1 & 2 & 3 & 4 & 5 \\ 
\midrule
$\b \delta_1$ & & 0.02 & 0.08 & 0.26 & 0.37 & 0.26 \\ 
\hdashline 
\multirow{5}{*}{$\b Q_1$}
& 1 & 0.07 & 0.76 & 0.17 & 0.00 & 0.00 \\ 
& 2   & 0.02 & 0.37 & 0.59 & 0.00 & 0.02 \\ 
& 3 & 0.00 & 0.01 & 0.55 & 0.43 & 0.02 \\ 
& 4 & 0.00 & 0.00 & 0.00 & 0.88 & 0.12 \\ 
& 5 & 0.00 & 0.00 & 0.00 & 0.00 & 1.00 \\ 
\midrule
$\b \delta_2$ && 0.12 & 0.37 & 0.34 & 0.13 & 0.04 \\ 

\hdashline 
\multirow{5}{*}{$\b Q_2$}
& 1 & 0.43 & 0.55 & 0.02 & 0.00 & 0.00 \\ 
& 2 & 0.17 & 0.70 & 0.13 & 0.00 & 0.00 \\ 
& 3 & 0.01 & 0.08 & 0.84 & 0.07 & 0.00 \\ 
& 4 & 0.00 & 0.00 & 0.00 & 0.99 & 0.01 \\ 
& 5 & 0.00 & 0.00 & 0.00 & 0.00 & 1.00 \\ 
\bottomrule
\bottomrule
\end{tabular}
\end{table}
As it can be noticed, the upper-level structure helps distinguish two different behaviors in the individual sequences. The first upper-level component identifies subjects who move quite rapidly towards higher hidden states; that is, it identifies participants whose cognitive functioning gets rapidly worse when compared to the study entrance. 
On the other hand, the second upper-level component denotes a moderate, albeit progressive, increase in the response variable over time, with individuals presenting a slower approach to more impaired statuses.

\subsection{Results: the missing data process}
\label{sec_LeidenMiss}
Maximum likelihood parameter estimates and the corresponding standard errors for the parameters in the missing data model are reported in Table \ref{tab_leidenMiss}. 

\begin{table}[!h]
\caption{Leiden 85+ data: parameter estimates and standard errors in the missing data model}
\label{tab_leidenMiss}

\centering
\begin{tabular}{lrrrrr}
\toprule
\toprule
 & \multicolumn{1}{c}{Estimates} & \multicolumn{1}{c}{Se} \\ 
\midrule
$\xi_1$ & -15.188 & 1.326 \\ 
$\xi_2$ & -8.809 & 0.782 \\ 
$\xi_3$ & -3.394 & 0.370 \\ 

\hdashline
$Age$& 2.474 & 0.268 \\ 
\textit{High Edu}& -1.628 & 0.519\\ 
$Male$ & 0.943 & 0.535\\ 
$\mbox{APOE}_{22-23}$ & 0.572 & 0.669\\ 
$\mbox{APOE}_{24}$& -0.351 & 0.637  \\ 
$\mbox{APOE}_{34-44}$ &  1.175 & 0.523\\ 

\bottomrule
\bottomrule
\end{tabular}
\end{table}
By looking at the estimated random intercepts, we may first observe that individuals  in the first class are less likely to drop-out from the study. The propensity to drop-out increases when we consider individuals belonging to the second class and, more heavily, to the third one. Generally, these tree latent classes may be labeled as completers, late, and early dropouts. 
As regards the fixed parameters, we may notice that both \textit{Age} and $\mbox{APOE}_{34-44}$ are positively associated with the probability of early exit, while such a probability reduces with higher education. As in the longitudinal data model, gender does not seem to substantially influence the missingness indicator.

Table \ref{tab_leidenClass} shows the estimated probabilities for the latent $U_i$, conditional on the $h$-th upper-level latent class, that is $ \pi_{k \mid h}$.
\begin{table}
\caption{Leiden 85+ data - missing data model: class probabilities $\b \pi_h$ by upper-level components}
\label{tab_leidenClass}
\centering
\begin{tabular}{lcccccc}
\toprule
\toprule
Class &\multicolumn{1}{c}{$\b \pi_1$}&\multicolumn{1}{c}{$\b 
\pi_2$}\\
\midrule
1 & 0.01 & 0.68 \\ 
2 & 0.42 & 0.17 \\ 
3 & 0.57 & 0.15 \\ 
\bottomrule
\bottomrule
\end{tabular}
\end{table}
Combining these results with those reported in Table \ref{tab_leidenMiss}, we may conclude that subjects in the first upper-level component present a higher chance to drop-out prematurely from the study after a premature and rapid decline of their cognitive functioning. This comes from an over-representation of the second and the third class in the first upper-level component of the the missing data model ($\pi_{2\mid 1} + \pi_{3\mid 1} = 0.99 $) which also corresponds to a non-persistent transition matrix in the longitudinal data model.

A step-by-step reduction of cognitive skills is instead observed for subjects who generally stay longer under observation and that belong to the second upper-level components.

\subsection{Sensitivity analysis}
Results discussed for the MMSE data are based on the assumption that the mechanism generating the observed data is MNAR. With the aim of verifying the sensitivity of parameter estimates to such an assumption, we present in this section the results obtained by the corresponding Missing At Random (MAR) model.
By looking at the BIC values in Table \ref{tab_leidenBic} obtained with $H = 1$, we may observe that the optimal MAR model corresponds to $G = 5$ and $K=3$; that is, no differences are present with respect to the MNAR counterpart when the number of states and classes is entailed. Table \ref{tab_leiden_sens} reports parameter estimates and standard errors for both the longitudinal and the missing data model.

\begin{table}[!h]
\caption{Leiden 85+ data: parameter estimates and standard errors in the longitudinal and the missing data model under the MAR assumption}
\label{tab_leiden_sens}

\centering

\begin{tabular}{lrrrrr}
\toprule
\toprule
& \multicolumn{2}{c}{Longitudinal} & \multicolumn{2}{c}{Missing }\\
\toprule
 & \multicolumn{1}{c}{Estimates} & \multicolumn{1}{c}{Se} &\multicolumn{1}{c}{Estimates} & \multicolumn{1}{c}{Se} \\ 
\midrule
$\zeta_1$ & 0.276 & 0.079 & - & - \\
$\zeta_2$ & 1.150 & 0.069 & - & - \\ 
$\zeta_3$ & 1.796 & 0.058 & - & - \\ 
$\zeta_4$ & 2.455 & 0.058 & - & - \\ 
$\zeta_5$ & 3.048 & 0.076 & - & - \\ 
\hdashline 
$\xi_1$ & - &- &  -16.086 & 1.863\\
$\xi_2$ & - &- & -9.187 & 1.017 \\ 
$\xi_3$ & - &- & -3.307 & 0.313 \\

\hdashline 
$Age$ & 0.052 & 0.008 & 2.764 & 0.401 \\ 
\textit{High Edu} & -0.280 & 0.071 & -2.172 & 0.756 \\ 
$Male$ & -0.083 & 0.070& 0.554 & 0.438  \\ 
$\mbox{APOE}_{22-23}$ &  0.044 & 0.055 & 0.474 & 0.408\\ 
$\mbox{APOE}_{24}$& -0.227 & 0.056 & -0.025 & 0.922\\ 
$\mbox{APOE}_{34-44}$& 0.245 & 0.087  & 0.800 & 0.438\\ 
\bottomrule
\bottomrule
\end{tabular}
\end{table}

When comparing the above results with those reported in Sections \ref{sec_LeidenLong}-\ref{sec_LeidenMiss}, we may not observe substantial differences, but for the estimate of $\mbox{APOE}_{34-44}$ in the missing data model that turns to be not significant under the MAR assumption. This may be possibly due to the reduced amount of information which can be exploited when the longitudinal and the missing data process are considered as independent. 
Apart from this difference, estimated parameters for the longitudinal data process seems to be quite robust to potential misspecification of the missing data mechanism and, therefore, render the results presented so far worth to be discussed. 

To complete the analysis, we also report in Table \ref{tab_MAR_Q} the estimated initial and transition probabilities for the longitudinal data model under the MAR assumption. 
\begin{table}
\caption{Leiden 85+ data - longitudinal data model: initial $(\b \delta)$ and transition ($\b Q$) probabilities under the MAR assumption}
\centering
\begin{tabular}{lcccccc}
\toprule
\toprule
 & State & 1 & 2 & 3 & 4 & 5 \\ 
\midrule
$\b \delta$ &  & 0.10 & 0.30 & 0.33 & 0.17 & 0.09 \\ 

\hdashline 
\multirow{5}{*}{$\b Q$}
& 1 & 0.42 & 0.57 & 0.02 & 0.00 & 0.00 \\ 
&  2 & 0.17 & 0.69 & 0.14 & 0.00 & 0.00 \\ 
&  3 & 0.01 & 0.07 & 0.82 & 0.09 & 0.01 \\ 
&  4 & 0.00 & 0.00 & 0.00 & 0.98 & 0.02 \\ 
&  5 & 0.00 & 0.00 & 0.00 & 0.00 & 1.00 \\ 
\bottomrule
\bottomrule
\end{tabular}
\label{tab_MAR_Q}
\end{table}
As it can be noticed, faster declines in the individual skills may not be recovered when ignoring the missingness process. Rather, the initial and transition probabilities which are estimated under such a modeling assumption are quite close to those associated to the second upper-level component in the MNAR counterpart (see Table \ref{tab_leiden_hmm}) which, as stated before, represents the most referenced category ($\tau_2=0.781$).

\section{Conclusions} \label{sec:5}
In this paper, we propose a a random coefficient based hidden Markov model for longitudinal responses subject to drop-out. We consider a non-homogeneous hidden Markov chain to capture unobserved dynamics in the longitudinal data model. Similarly, a non-homogeneous finite mixture is considered for the missing data model to capture sources of unobserved heterogeneity that influence premature exits from the study. Last, an upper-level mixture is introduced to model the dependence between the random coefficients in the two profiles of interest.
The proposed model specification offers great flexibility and provides a clear and concise description of the observed data. The application to data from the Leiden 85+ study shows the strengths of the model and leads to the identification of two well distinguished sets of study participants. The former is characterized by subjects who drop-out out prematurely from the study after experiencing a severe health worsening, as it is clear from the estimated transition probability matrix. On the other hand, the second upper-level class identifies subjects who stay longer under observation experiencing moderate, albeit progressive, health worsening during the follow-up. 
As regards the effect of socio-demographic and genetic factors on the evolution of cognitive functioning in the elderly, the results discussed so far are in line with previous studies on the topic. As expected, age negatively influences individual skills, a higher educational level represents quite a protective factor, while $\epsilon_4$ carriers seems to present a slower decline of their cognitive functioning.

\end{document}